\documentclass[superscriptaddress,aps,prd,twocolumn,preprintnumbers,nofootinbib,reprint]{revtex4-2}
\usepackage[utf8]{inputenc}
\usepackage{uniinput}
\usepackage{amsmath,mathtools,mathrsfs,hyperref}
\hypersetup{
	colorlinks=true,
	allcolors=blue
}

\usepackage{microtype}
\usepackage[plain]{fancyref}
\usepackage{enumitem,subfigure}

\renewcommand{\d}{\mathrm{d}}
\newcommand{\ie}{\textit{i.e.}}

\newcommand{\ttt}{\tilde{t}}
\newcommand{\xt}{\tilde{x}}
\newcommand{\yt}{\tilde{y}}
\newcommand{\phit}{\tilde{ϕ}}
\newcommand{\tin}{t_{\textrm{in}}}
\newcommand{\tout}{t_{\textrm{out}}}
\newcommand{\yin}{y_{\textrm{in}}}
\newcommand{\yout}{y_{\textrm{out}}}
\newcommand{\phiin}{ϕ_{\textrm{in}}}
\newcommand{\phiout}{ϕ_{\textrm{out}}}

\begin{document}
	
	\title{Horizonless black hole mimickers with spin}
	
	\author{Ulf Danielsson~}
	\email{ulf.danielsson@physics.uu.se}
	\affiliation{Institutionen för fysik och astronomi,
		Uppsala Universitet, Box 803, SE-751 08 Uppsala, Sweden}
	\author{Suvendu Giri~}
	\email{suvendu.giri@princeton.edu}
	\affiliation{Institutionen för fysik och astronomi,
		Uppsala Universitet, Box 803, SE-751 08 Uppsala, Sweden}
	\affiliation{Department of Physics, Princeton University, Princeton, New Jersey 08544, USA}
	\affiliation{Princeton Gravity Initiative, Princeton University, Princeton, NJ 08544, USA}

	\preprint{UUITP-29/23}

	\begin{abstract}
	\noindent
	In this paper, we study rotating horizon less black shells as an alternative to Kerr black holes. We make use of Ernst's potential to solve the Einstein equations perturbatively in the angular momentum $a$. Calculating to order $a^6$, we find accurate predictions up to about $a=0.45$, where the quadrupole moment is predicted to be around $1\%$ higher than its Kerr value; higher multipole moments show deviation of the order of $\sim 10\%$. Our analysis takes into account deformations of the black shells, and we propose that it can be used for numerical simulations comparing gravitational waves emitted by orbiting black shells with those emitted by orbiting black holes. 
    We find that matter on the rotating shell takes the form of a viscous fluid. We make extensive use of relativistic hydrodynamics, and discover an intricate structure of circulating flows of this fluid and heat on the  black shell, sustained by the Unruh effect.
	\end{abstract}
	
	\maketitle
	\tableofcontents

\allowdisplaybreaks

\section{Introduction and summary}\label{sec:intro}
	
Given the profound difficulties in formulating a consistent quantum version of a black hole, it is interesting to study alternative scenarios, where genuine black holes do not exist. In \cite{Danielsson:2017riq} we proposed such an alternative that makes use of the building blocks available in string theory. The key idea is that the vacuum is unstable against nucleation of bubbles of true vacuum with a negative cosmological constant. This is natural in string theory, where there are such anti de Sitter (AdS) vacua in abundance. Under normal conditions, the nucleation is heavily suppressed, consistent with the long term stability of our vacuum. However, when matter threatens to collapse into a black hole a new possibility opens up in phase space. At the same time as the bubble forms, the infalling matter is converted into a gas of massless open strings attached to the brane that constitutes the bubble wall. Again, this is a natural construction from the point of view of string theory. The gas will be kept at the local Unruh temperature, and carries an entropy of the order of the corresponding black hole. This completely overwhelms the suppression against tunneling, and instead of being highly unlikely, the nucleation becomes inevitable.
	
In \cite{Danielsson:2017riq} it was argued that the natural radius for the resulting bubble will be the Buchdahl radius at $R_{\textrm{\scshape b}}=9R_{\textrm{\scshape s}} /8$. The reason was that the energy momentum tensor on the shell, induced by the extrinsic curvature, takes the form of a massless gas at this radius. It is well known that such gravastar like constructions (usually using an interior of de Sitter(dS) rather then AdS as in our case) are inherently unstable. The solution, already proposed in \cite{Danielsson:2017riq}, was that there should exist an interaction term that allows the transfer of energy between the different matter components of the model. The tension of the brane is supposed to be close to its critical value, but the idea is that it can release energy to the gas, and absorb it again, in such a self controlled way that stability can be achieved. This mechanism was studied in detail in \cite{Danielsson:2021ykm}, where it was shown to work with parameters within a certain range. The criteria for stability were that vibrations around the equilibrium should be damped, and that matter accreting on the brane should yield a shell that grows in radius and relaxes to the new Buchdahl radius. That this actually is possible is a highly non-trivial test of the proposal.
	
In \cite{Danielsson:2017pvl,Danielsson:2021ruf} we tried to generalize the construction to spinning black holes with a small angular momentum. This is important if we want to identify a signal that can be used to observationally distinguish a black shell from a black hole. Note that, as argued in \cite{Danielsson:2017pvl}, the shell is as effective in absorbing matter as a genuine black hole. This is due to its energy being carried by a gas at an extremely low temperature (almost as low as the Hawking temperature), and consequently very high entropy. It is only through Hawking scale thermal radiation that energy can be released. Hence, the black shell is as black as a black hole. The absence of a true horizon makes a more profound difference if the black shell is rotating. The uniqueness of the Kerr metric, given the mass and the angular momentum of the black hole, depends on the requirement that the horizon is non-singular. If the spacetime is cut off before the horizon, there is no longer any restriction on the higher multipole moments. In case of the Earth, for instance, the multipole moments of the surrounding spacetime, depend on the detailed shape of the Earth and its distribution of mass. As a curiosity, one might note that the angular momentum of the Earth is three orders of magnitude larger than the maximum value allowed for a Kerr metric with the same mass.
	
In \cite{Danielsson:2017pvl,Danielsson:2021ruf} we assumed that the matter on top of the brane was a perfect fluid. This led us to a prediction of a specific change in the quadrupole moment compared to Kerr. The coordinates used in \cite{Danielsson:2017pvl,Danielsson:2021ruf}, were, however not well suited for calculations beyond the lowest order in the angular momentum. In this paper we introduce a much better set of coordinates which we use to extend the analysis in a systematic way to arbitrary order in the angular momentum. In doing so we also discover that the condition of having a perfect fluid is inconsistent with the condition of a traceless energy momentum tensor. We therefore reverse the order of applying the constraints, and start out by imposing a condition of traceless energy momentum tensor to identify the gas on top of the brane that the infalling matter gives rise to. Interestingly, this constraint uniquely fixes the size and shape of the shell together with the full external space time. On the other hand, we also discover that \emph{the fluid is far from perfect}. Its physical properties are much more interesting with \emph{finite heat conductivity} as a well as \emph{non-zero shear viscosity}. The latter, as we will discuss, is intimately connected to the black shell being an almost perfect absorber.\footnote{It has recently been claimed \cite{Carballo-Rubio:2022imz} that observations by EHT prohibit compact objects that are larger than horizon size. However, to reach this conclusion one must assume an absorption coefficient close to zero. For good absorbers like our black shell, there is no meaningful bound on compactness.}
	
The outline of the paper is as follows. In \fref{sec:rotating}, we review our previous results formulated using the new coordinates, clearly stating the assumptions that go into our choice of solutions. We then turn up the spin and consider contributions at higher orders, discovering the need to consider a non-perfect fluid. Our results yield useful predictions for the quadrupole up to a spin of $a=0.45m$, and the analysis can be extended to higher orders.  In \fref{sec:hydrodynamics}, we study the hydrodynamics of the model obtaining a very satisfactory picture of the flow of fluid and heat on the rotating body. Amazingly, the picture that we find is very similar to the trade winds blowing on the Earth. In \fref{sec:stability} we present an argument as to why the black shell is stable against small perturbations. In the final section we briefly discuss the observational consequences.

\section{Rotating black shells}\label{sec:rotating}
	
The stationary black shell, as reviewed in the introduction, is described by Israel's junction conditions across the shell. To find our solution we need to make a number of choices based on physical intuition. 
\begin{enumerate}[label=(\roman*)]
    \item We assume that the nucleated piece of AdS-space is not rotating. The intuition behind the assumption is that the fresh piece of space appearing through the tunneling should not have any prior knowledge of the surrounding space time that it will attach to.
    \item We assume that the bubble wall and the matter attached to it contains radiation with a traceless energy momentum tensor.\footnote{In \cite{Danielsson:2021ruf} we started out by demanding a perfect fluid. This, it turns out, is incompatible with a trace-less fluid at higher orders. We are therefore forced to relax the condition that the fluid should be perfect.} Here we define the energy momentum tensor by performing a subtraction of the empty space. Remarkably, this {\it uniquely} fixes the exterior space time, as well as the radius and shape of the shell.  The exterior will be a modification of the Kerr spacetime, fully specified by its multiple moments.
\end{enumerate}
Amusingly, there is a very similar story going all the way back to work by Einstein in 1913, as well as by Lense and Thirring in \cite{thirring1918effect,Lense:1918zz,Mashhoon:1984fj}. The goal at the time was to see if Mach's principle can be realized inside of a rotating shell of matter. Already in these early works, it was clear that space-time was dragged along by the rotating shell at least to some extent. However, it was not until the mid 1980s \cite{Pfister_1985,Pfister_1986} that the problem was fully solved. In \cite{Pfister_1985} it was shown that a piece of flat space time can be glued on as the interior of a thin shell that is rotating compared to an exterior, asymptotic observer. Inside of the shell, space-time is {\it exactly} Minkowski without any Coriolis forces. An observer inside of the shell would not notice any rotation, but looking outwards would discover the presence of the rotating shell surrounded by a rotating universe. This is how far Mach's principle is implemented in general relativity. As shown in \cite{Pfister_1985}, the exterior metric and its multipole moments are unique up to the radius of the shell if one imposes perfect fluid. See \cite{Pfister:2015ftd} for an illuminating recent review.

Our case is different in two ways. First, the interior of the bubble is AdS (again without any effects from rotation) instead of flat space, where we work in the limit of $r \gg R_{\textrm{AdS}}$. On the other hand, to identify the gas attached to the brane we first make a subtraction, where we compare to a bubble in flat space. In practice, this first step formally involves the junction between the exterior rotating solution and that of non-rotating flat space. This would be just the same as \cite{Pfister_1985} if we were to impose perfect fluid. Instead, we demand that the fluid has a traceless energy momentum tensor. This is actually a stronger condition that completely fixes everything, even though the fluid is not perfect. 

After this general description of our strategy, let us move on to solving the Einstein equations and the junction conditions.

\subsection{Rotating metrics using the method by Ernst}
	
The most general stationary axially symmetric solution of vacuum Einstein equations is given by the Weyl-Lewis-Papapetrou line element, which in prolate spheroidal coordinates (with $x\geq 1,-1 \leq y \leq 1$) reads
\begin{equation}\label{eq:wlp}
\begin{aligned}
	\d s² &= -f (\d t - ω \d ϕ)² + \frac{σ²}{f} \left[ e^{2γ}\left( x²-y² \right) \phantom{\frac{\d x²}{x²-1}} \right. \\
	& \left. \left( \frac{\d x²}{x²-1} +\frac{\d y²}{1-y²} \right) + \left( x²-1 \right) \left( 1-y² \right)\d ϕ² \right],
\end{aligned}
\end{equation}
where the functions $f\equiv f(x,y),ω \equiv ω(x,y), γ \equiv γ(x,y)$ depend on coordinates $(x,y)$ only, and $σ\coloneqq \sqrt{m²-a²}$.
Using a scalar $Ω$ defined in terms of the metric functions $f$ and $ω$
\begin{equation}
	σ (x²-1) ∂_x Ω = f² ∂_y ω\,, \qquad
	σ (1-y²) ∂_y Ω = -f² ∂_x ω\,,
\end{equation}
we can now define a complex function (Ernst's potential)
\begin{equation}
	ξ = \frac{1-E}{1+E}\,,\quad
	\text{with }\quad
	E = f + i Ω\,.
\end{equation}
Einstein's equations for $f$ and $ω$ can be recast into a compact form (called Ernst's equation) in terms of this potential
\begin{equation}\label{eq:Ernst_eq}
	\left(ξ ξ^* - 1\right) ∇²ξ =2ξ^* ∇ξ \cdot ∇ξ \,,
\end{equation}
where $∇$ and $∇²$ are two-dimensional gradient and Laplacian in $(x,y)$ coordinates i.e.
\begin{equation}
\begin{aligned}
	&\left(ξ ξ^* - 1 \right) 
	\left\{
	∂_x\left[ \left(x²-1\right)∂_xξ 
	\right] +
	∂_y\left[ \left(1-y²\right)∂_yξ 
	\right]\right\}\\
	&\qquad =
	2ξ^* 
	\left[ \left(x²-1\right)∂_{xx}ξ + \left(1-y²\right)∂_{yy}ξ
	\right]\,.
\end{aligned}
\end{equation}
$ξ$ is symmetric under the exchange $x \leftrightarrow y$. The simplest solution to Ernst's equation is $ξ^{-1} = x$, which gives the Scwharzschild spacetime. Symmetry of the Ernst potential then dictates that $ξ^{-1}=y$ is also a solution. The linear combination $ξ^{-1} = (σ/m) (x + i y)$ also solves Ernst's equation and corresponds to  the Kerr solution, which is explicitly given by
\begin{align}
	f &= \frac{σ²x²+a²y²-m²}{\left(σx+m\right)²+a²y²},\nonumber\\
	ω &= 2am\frac{\left(σx+m\right)\left(1-y²\right)}{σ²x²+a²y²-m²},\\
	e^{2γ} &= \frac{σ²x²+a²y²-m²}{σ²\left(x²-y²\right)}\,.\nonumber
\end{align}
For $a=0$, this reduces to the Schwarzschild solution
\begin{equation}
	f = \frac{x-1}{x+1},\quad
	ω = 0,\quad
	e^{2γ} = \frac{x²-1}{x²-y²}\,.
\end{equation}
Prolate spheroidal and Boyer-Lindqvist coordinates are related by the transformation
\begin{equation}\label{eq:PS_BL}
	x = \frac{r-m}{σ},\quad y =\cosθ\,.
\end{equation}
Starting from the general static and axisymmetric solution in \eqref{eq:wlp}, Quevedo and Mashhoon \cite{Quevedo:1989rfm,Quevedo:1991zz} found a general solution by introducing an arbitrary Zipoy-Voorhees (ZV) parameter $δ$, and applying Hoenselaers-Kinnersley-Xanthopoulos transformations \cite{HKX} which introduces an infinite number of multipole moments. We will refer to this solution as the (Quevedo-Mashhoon) QM solution from hereon. Explicitly, the solution reads
\begin{align}
		f &= \frac{R}{L} e^{-2δ \hat{ψ}}\,,\nonumber\\
		ω &= 2a - 2 σ \frac{\mathcal{M}}{R}e^{2δ \hat{ψ}}\,,\\
		e^{2γ} &= \frac{1}{4}\left( 1+\frac{m}{σ} \right)² \frac{R}{\left( x²-1 \right)^δ} e^{2δ² \hat{γ}}\,,\nonumber
\end{align}
where the quantities involved are defined below
\begin{align}\label{eq:sub1}
    	R &= a_+ a_- + b_+ b_-\,,\nonumber\\
		L &= a_+² + b_+²\,,\nonumber\\
		\hat{ψ} &= \sum_{n=1}^{∞}q_n Q_n P_n\,,\\
		\mathcal{M} &= (x+1)^{δ-1} \left[ x(1-y²)(λ+μ) a_+ + y(x²-1)(1-λ μ)b_+ \right] \,,\nonumber\\
		\hat{γ} &= \sum_{m,n=0}^∞ (-1)^{m+n} q_m q_n \int_{-1}^y Γ_{m,n}\,.\nonumber
\end{align}
These are further defined in terms of
\begin{align}\label{eq:sub2}
		a_\pm &= (x\pm1)^{δ-1} \left[ x(1-λ^+ λ^-) \pm (1+λ^+ λ^-) \right]\,,\nonumber\\
		b_\pm &= (x\pm1)^{δ-1} \left[  y(λ^+ + λ^-)\, \pm (λ^+ - λ^-)  \right]\,,\nonumber\\
		λ^\pm &= α (x²-1)^{1-δ} (x\mp y)^{2(δ-1)}\nonumber\\
		&\quad  \exp \left[2δ \sum_{n=1}^∞ (-1)^n q_n β_n^\pm\right]\,,\nonumber\\
		β_n^\pm &= (\pm 1)^n\left(\frac{1}{2}\ln \frac{(x\mp y)²}{x²-1} - Q_1\right) + P_n Q_{n-1}\\ 
		& \quad 
		- \sum_{k=1}^{n-1} (\pm 1)^k P_{n-k} \left( Q_{n-k+1} - Q_{n-k-1} \right)\,,\nonumber\\
		Γ_{m,n} &= \left(x²-1\right)P_m^\prime Q_m^\prime \left(2x P_n Q_n^\prime - y P_n^\prime Q_n\right)\nonumber\\
		&\quad + \frac{\left(x²-1\right)²}{x²-y²} \left[ P_m Q_m^\prime \left( y P_n Q_n^\prime - x P_n^\prime Q_n \right) \right.\nonumber\\
		& \quad \left. + P_m^\prime Q_m \left( y P_n^\prime Q_n - x P_n Q_n^\prime \right)
		\right]\,.\nonumber
\end{align}
In the above, $q_n$ with $n = 0,1,2,\ldots$ are constant parameters.
$P_n \equiv P_n(y), Q_n \equiv Q_n(x)$, are Legendre polynomials of the first and second kind (in the domain $x²\geq 1, y² \leq 1$), and $P_n^\prime, Q_n^\prime$ are their derivatives respectively. Explicitly,
\begin{align}
	P_0 &=1, P_1 = y, P_2 = \frac{1}{2} \left(3y²-1\right), \, \ldots \nonumber\\
	Q_0 &= \frac{1}{2} \ln \frac{x+1}{x-1},
	Q_1 = x Q_0 -1,\\ \nonumber
	Q_2 &= \frac{1}{2} \left(3x²-1 \right)Q_0 - \frac{3x}{2}, \, \ldots\\
	P_n^\prime &\coloneqq ∂_y P_n, \quad 
	Q_n^\prime \coloneqq ∂_x Q_n\,. \nonumber
\end{align}
Before even making an ansatz for solving the junction conditions, it is useful to take a look at the multipole moments. To study the multipolar structure of this spacetime, we can compute the Geroch-Hansen multipole moments using the prescription by Fodor-Hoenselaers-Perjés \cite{FHP} which uses Ernst's potential. We choose the parameters $q_0=1, q_{2n} \neq 0, q_{2n+1} = 0$. Below we list the first few multipole moments for arbitrary $δ,q_n$.
\begin{align}\label{eq:multipoles}
	M_0 &= m+(\delta -1) \sigma \,,\nonumber\\
	M_2 &= -m^3-3 (\delta -1) m^2 \sigma -(\delta -2) \delta  m \sigma ^2 \nonumber \\
	& +\sigma ^3 \left[-\frac{\delta
   ^3}{3}+\delta ^2+\frac{2}{15} \delta  (q_2+10)-2\right]\,,\nonumber\\
	M_4 &=m^5+5 (\delta -1) m^4 \sigma +2 \left(3 \delta ^2-6 \delta +2\right) m^3 \sigma ^2\nonumber\\
    & +\frac{2}{105} m^2
   \sigma ^3 \left[ \hspace*{-7pt}\phantom{\frac{1}{1}} 115 \delta ^3-345 \delta ^2-2 \delta  (11 q_2+35) \right. \nonumber \\
   & \left. +300 \hspace*{-7pt}\phantom{\frac{1}{1}}\right] +\frac{1}{105} m
   \sigma ^4 \left[95 \delta ^4-380 \delta ^3 \right. \nonumber \\
   & \left. +\delta ^2 (10-32 q_2)+4 \delta  (8  q_2+185)-360  \hspace*{-7pt}\phantom{\frac{1}{1}} \right]\nonumber\\
   & +\frac{1}{315} \sigma ^5 \left[ \hspace*{-7pt}\phantom{\frac{1}{1}} 57 \delta ^5-285 \delta ^4-6 \delta ^3 (8
   q_2-55) \right. \nonumber \\
   & \left. \hspace*{-7pt}\phantom{\frac{1}{1}} +6 \delta ^2 (16 q_2+25)+\delta  (72 q_2+8 q_4+468)-720\right]\,, \nonumber \\
   M_6 &= -m^7-7 (\delta -1) m^6 \sigma -3 \left(5 \delta ^2-10 \delta +4\right) m^5 \sigma ^2\nonumber\\
   &  -\frac{1}{63} m^4
   \sigma ^3 \left[775 \delta ^3-2325 \delta ^2+\delta  (1232-46 q_2)+318\right]\nonumber\\
   & -\frac{1}{105} m^3
   \sigma ^4 \left[ \hspace*{-7pt}\phantom{\frac{1}{1}} 625 \delta ^4-2500 \delta ^3+\delta ^2 (970-176 q_2) \right. \nonumber \\
   & \left.  \hspace*{-10pt}\phantom{\frac{1}{1}} +4 \delta  (44
   q_2+765)-1840\right] -\frac{m \sigma ^6}{3465}  \left[  \hspace*{-7pt}\phantom{\frac{1}{1}} 2723 \delta ^6\right. \nonumber \\
   & -16338 \delta ^5  +\delta ^4 (27510-2024 q_2) \nonumber \\
   & +8 \delta ^3 (759 q_2-140)+8 \delta ^2 \left(6 q_2^2-73 q_2+34 q_4+29\right)\nonumber \\
   &\left. -8 \delta  (433 q_2+34 q_4+5224)+25320  \hspace*{-7pt}\phantom{\frac{1}{1}}\right]\nonumber \\ 
   & -\frac{m^2 \sigma ^5}{3465}\left[7623 \delta ^5  \hspace*{-7pt}\phantom{\frac{1}{1}} -38115 \delta ^4+\delta ^3
   (33610-3456 q_2) \right. \nonumber\\
     & +6 \delta ^2 (1152 q_2+8605) +4 \delta  (284 q_2+78
   q_4-10687)\nonumber\\
   &\left. -12000  \hspace*{-7pt}\phantom{\frac{1}{1}} \right] +\frac{\sigma ^7 }{45045}\left[  \hspace*{-7pt}\phantom{\frac{1}{1}} -5057
   \delta ^7 +35399 \delta ^6\right. \nonumber \\
   &+26 \delta ^5 (253 q_2-3725)-52 \delta ^4 (506 q_2-2505)\nonumber\\
   &-52 \delta ^3  \left(12 q_2^2-441 q_2+34 q_4+2189\right) \nonumber \\
   &+52 \delta ^2 \left(12 q_2^2+130 q_2+68  q_4+1993\right)\nonumber\\
   &  \left.+40 \delta  (416 q_2+52 q_4+6 q_6+1625)-118560  \hspace*{-7pt}\phantom{\frac{1}{1}} \right]\,,\nonumber\\
	& \vdots \\
	J_1 &= a m + 2 a (\delta -1) \sigma\,,\nonumber\\
	J_3 &=-3 a m^3 (\delta -1)^2 +a^3 m\left(3 \delta ^2-6 \delta +2\right)\nonumber\\
    &-4 a (\delta -1) m^2 \sigma\nonumber\\
   & +\frac{2}{15} a \sigma ^3 \left(-5 \delta ^3+15 \delta ^2+\delta  (2 q_2+5)-15\right)\,,\nonumber\\
	J_5 &= a m^5+6 a (\delta -1) m^4 \sigma\nonumber\\ 
    &+2 a \left(5 \delta ^2-10 \delta +4\right) m^3 \sigma ^2\nonumber\\
    &  +\frac{4}{7}
   a m^2 \sigma ^3 \left(10 \delta ^3-30 \delta ^2-\delta  (q_2-14)+6\right)\nonumber\\
   &  +\frac{1}{63} a m \sigma
   ^4 \left[  \hspace*{-7pt}\phantom{\frac{1}{1}} 145 \delta ^4-580 \delta ^3+\delta ^2 (350-52 q_2) \right. \nonumber \\
  & \left. +\delta  (52 q_2+460)-312  \hspace*{-7pt}\phantom{\frac{1}{1}} \right] +\frac{2}{315} a \sigma ^5 \left[  \hspace*{-7pt}\phantom{\frac{1}{1}} 37 \delta ^5 \right. \nonumber \\
  & -185 \delta ^4 -40 \delta ^3
   (q_2-6)+20 \delta ^2 (4 q_2+1)\nonumber \\ 
   & \left. +\delta  (38 q_2+8 q_4+278)-390  \hspace*{-7pt}\phantom{\frac{1}{1}} \right]\,,\nonumber\\
	& \vdots \nonumber\\
	&M_{2k+1} = J_{2k} = 0,\quad k=0,1,2,\ldots\nonumber
\end{align}
The odd mass multipole moments and even current multipole moments vanish because of the equatorial symmetry of the spacetime ($y \leftrightarrow -y$).
In the limit $δ=1$ and $q_{n} =0$, the QM metric reduces to Kerr, and so do the multipole moments. It is also interesting to note that in the extremal limit $a → m ⇒ σ → 0$, the multipoles reduce to those of extremal Kerr regardless of the values of the parameters $\delta$ and $q_{2n}$.
	
The QM metric is asymptotically flat and smooth outside the sphere $x=1$ which is a naked singularity. It is therefore a good description of the spacetime outside an axisymmetric rotating object. Our shell sits outside this singular surface. We will therefore use this metric to describe the spacetime outside the rotating black shell. To find a complete description of the internal structure of the black shell, we will use Israel's junction conditions to find the energy momentum tensor on the shell that can support the AdS vacuum in its interior.

Spacetime inside the shell is AdS (with cosmological constant $Λ=-3k²$) which, in prolate spheroidal coordinates, can be written as 
\begin{align}
	\d s²_{\textrm{AdS}} &=
	-\frac{\xt+1}{\xt-1} \d \ttt\,² +
	\frac{\d \xt²}{2 k^2 (\xt-1)^2 (\xt+1)}\nonumber\\
	& + \frac{2 \d \yt²}{k^2 (\xt-1) \left(1-\yt^2\right)} + 
	\d ϕ² \frac{2 \left(1-\yt^2\right)}{k^2 (\xt-1)}\,,
\end{align}
This is related to global coordinates by the change of coordinates 
\begin{equation}\label{eq:PS_AdS}
    \xt = 1 + 2 k^{-2}\tilde{r}^{-2},\quad \yt = \cos \tilde{θ}.
\end{equation}
Before proceeding, let us make a quick comment on how the QM metric relates to the exterior metric in \cite{Danielsson:2021ruf}. For a choice of parameters, the exterior spacetime there can be chosen to be the Hartle-Thorne (HT) metric \cite{Hartle:1968si}, which is an approximate solution to vacuum Einstein's equations to order $a²$, and represents the spacetime outside a slowly rotating object. The QM metric reduces to the HT metric for $δ=1, q_0=1, q_2 \neq 0, q_{n\geq 4}=0$. Being an exact solution of Einstein's equations in vacuum, the QM metric can be thought of as an extension of the HT metric to all orders in spin. Since we intend to extend our results in \cite{Danielsson:2021ruf} (which were to order $a^2$) to all orders in spin, the QM metric is indeed a natural choice. This metric has been well studied, and its properties well understood. See \cite{Bini:2009cg} for a recent discussion.

\subsection{Junction conditions}
Having specified the spacetimes inside and outside the bubble in their respective coordinate systems: $\left(t,x,y,ϕ\right)$ outside and $\left(\ttt,\xt,\yt,\phit\right)$ inside, let us now proceed to solve Israel's junction conditions on the shell which involves:
\begin{enumerate}[label=(\roman*)]
	\item \emph{first junction condition}: ensuring continuity of the induced metric across the shell
	\item \emph{second junction condition}: computing the energy-momentum tensor on the shell ($S^a{}_b$) from the difference in the extrinsic curvatures ($K^a{}_b$) across the shell
	\begin{equation}\label{eq:jc2}
		-8πS^a{}_b = ΔK^a{}_b - ΔK δ^a{}_b\,.
	\end{equation}
\end{enumerate}
To do this, we first need to pick a coordinate system on the shell. For simplicity, let us use the coordinates from the outside: $\left(t,y,ϕ\right)$. We will parameterize the position of the shell as an expansion in spin, with the zeroth order coefficients given by our $a=0$ result from \cite{Danielsson:2017riq}. We choose this expansion only for ease of computation. The bulk metrics are exact vacuum solutions of Einstein's equations, so it is possible to extend the solution to all orders in $a$. For simplicity, we will restrict ourselves to an expansion to order $a^6$ in this paper and treat the problem perturbatively order by order in $a$. We take the embedding of the shell in the coordinate system at infinity to be
\begin{equation}\label{eq:paramout}
	x = \frac{5}{4}
	+  \sum_{m=1}^3 a^{2m} \sum_{n=0}^m x_{2m,2n}y^{2n}\,.
\end{equation}
With respect to the coordinate system at the center of the bubble, we take the embedding to be
\begin{align}\label{eq:paramin}
	\xt &= \left( 1+\frac{32}{81k²m²} \right) 
	+ \sum_{m=1}^3 a^{2m} \sum_{n=0}^m \xt_{2m,2n}y^{2n} \,,\nonumber\\
	&\yt =y \,,
	\phit = ϕ\,,
	\ttt = \frac{4t}{\sqrt{16+81k²m²}}\,.
\end{align}
Zeroth order position of the shell $x$ and $\xt$ are simply the Buchdahl radius $r=9m/4$ in the respective coordinate systems obtained using \eqref{eq:PS_BL} and \eqref{eq:PS_AdS} respectively.
We will make two further simplifying assumptions on the exterior spacetime which ensure that the multipole moments start receiving corrections at the same order as Kerr, and the corrections vanish when $q_n=0$.
\begin{enumerate}[label=(\roman*)]
	\item we will restrict ourselves to a constrained class of the QM solution where the ZV parameter $δ$ and $q_n$ are related via
	\begin{equation}
		δ^{-1} = \sum_{n=0}^{∞} q_n  = 1 + \sum_{k=1}^{∞} q_{_{2k}} \,.
	\end{equation}
	\item We assume that the parameters $q_{n}$ start contributing at order $a^n$ in spin and have an expansion of the form
	\begin{equation}
		q_{_{2k}} = \sum_{i=0}^{∞} q_{_{2k,2i}}\, a^{2(k+i)}\,.
	\end{equation}
\end{enumerate}
With these assumptions, the solution for all $a$ will be uniquely fixed by having the non-rotating black shell with $a=0$ to sit at the Buchdahl radius.

\subsection{Solving order by order in spin}
To solve the first junction condition, we compute the metric induced on the shell from both sides and demand that they are equal up to a coordinate transformation of the form\,\footnote{This is equivalent to the freedom to choosing a different parametrization of the shell as seen from inside in \eqref{eq:paramin}. The reason for this choice is purely the technical ease of solving the equations. It should be possible to reformulate the analysis to eliminate this, but since it makes no difference to the result and is much easier to work with, for the purpose of this already tedious computation, we have chosen this route.}
\begin{equation}\label{eq:coord_jc1}
	\left\{
	\begin{aligned}
		\tin &= \tout + \sum_{m=1}^3 a^{2m} g_{2m}\left(\tout,\phiout\right)\,,\\
		\yin &= \yout + \sum_{m=1}^3 a^{2m} y_{2m}\left(\yout\right) \,,\\
		\phiin &= \phiout + \sum_{m=0}^2 a^{2m+1} f_{2m}\left(\tout,\phiout\right) \,.
	\end{aligned} \right.
\end{equation}
To order $a²$, this determines the functions in \eqref{eq:coord_jc1}
\begin{align}
		\frac{g_2}{\tout} &= \frac{3 (41 \ln 3-20) \left(81 k^2 m^2+8\right)}{32 \left(81 k^2 m^2+16\right)}q_{2,2}\nonumber \\
		& +\frac{\left(972 k^2 m^2+256\right)}{729 k^2 m^2+144}x_1 -\frac{37503 k^2 m^2+9536}{1458 m^2 \left(81 k^2 m^2+16\right)}\,,\nonumber\\
		y_2 &= \yout \left(1-\yout²\right) \frac{\left(2187 m^2 q_{2,2} (5 \ln 3-4)-544\right)}{2916 m^2},\nonumber \\
		f_2 &= \frac{128}{729m²} \tout \,,
\end{align}
and the embedding parameters
\begin{align}\label{eq:jc1_sol_x}
		\xt_{2,0} &= \frac{81 m^2 (27 q_{2,2} (41 \ln 3-20)-256 x_1)+4256}{59049 k^2 m^4}\,,\nonumber\\
		\xt_{2,2} &= -\frac{\left(81 k^2 m^2+16\right) \left(81 m^2 q_{2,2} (615 \ln 3-556)-2048\right)}{2187 k^2 m^4 \left(243 k^2 m^2+64\right)}\,,\nonumber\\
		x_2 &= \frac{243 m^2 k^2 \left(729 m^2 q_{2,2} (99 \ln 3-92)-2816\right)}{10368 m^2 \left(243 k^2 m^2+64\right)} \\
		& + \frac{17496 m^2 q_{2,2} (59 \ln 3-60)-32768}{10368 m^2 \left(243 k^2 m^2+64\right)}\nonumber
\end{align}
This can be done order by order in $a$ to arbitrary orders. Since the higher order expressions are not particularly illuminating, instead of reproducing them here, we have included all results up to order $a^6$ in a Wolfram Mathematica notebook \cite{mathematica:a6}.\footnote{It is straightforward to extend this beyond $a^6$ with the only trade-off being that the metric and higher corrections become increasingly complicated to work with. We stop at $a^6$ in this paper as this is enough to study qualitative properties of the black shell, as well as make accurate predictions to spin $a^6 \sim 1 \% ⇒ a \sim 0.45m$.}
	
Next, we compute the extrinsic curvature of the shell $K^a{}_b$ from both sides using the embedding in \eqref{eq:paramout} and \eqref{eq:paramin}. To use Israel's junction condition \eqref{eq:jc2} we need them both in the same coordinate system. So we transform $K_{\textrm{in}}$ using the coordinate transformation \eqref{eq:coord_jc1}. Now, we can finally compute the stress tensor on the shell $S^a{}_b$. The full expression is complicated, so below we present it in the large $k$ limit that is relevant to the problem at hand. The full stress tensor to order $a^6$ can be found in the Wolfram Mathematica notebook \cite{mathematica:a6}.
	\begin{align}
		S^t{}_t &= -\frac{k}{4 \pi }+\frac{1}{27 \pi  m}\nonumber\\
		& +a^2 \frac{27 m^2 (q_{2,2}(5724-5859 \ln 3)+256 x_1)-736}{139968 \pi  m^3} \nonumber\\
		& +a²y^2 \frac{ \left(2187 m^2 q_{2,2}(733 \ln 3-676)-40960\right)}{419904 \pi  m^3}\,,\nonumber\\
		S^y{}_y &= -\frac{k}{4 \pi }+\frac{5}{54 \pi  m}\nonumber\\
		& + a^2 \frac{m^2 (135 q_{2,2}(44-39 \ln 3)-5376 x_1)+800}{31104 \pi  m^3}\nonumber\\
		&-a²y^2 \frac{\left(729 m^2 q_{2,2}(93 \ln 3-164)+2048\right)}{839808 \pi  m^3}\,,\nonumber\\
		S^ϕ{}_ϕ &= -\frac{k}{4 \pi }+\frac{5}{54 \pi  m}\\
		& + a^2 \frac{10592-9 m^2 (9 q_{2,2}(1239 \ln 3-1004)+1792 x_1)}{93312 \pi  m^3}\nonumber\\
		&+ a²y²\frac{\left(729 m^2 q_{2,2}(951 \ln 3-620)-75776\right)}{839808 \pi  m^3}\,,\nonumber\\
		S^t{}_ϕ &= \frac{2 a \left(y^2-1\right)}{9 \pi  m},\quad
		S^ϕ{}_t = \frac{32 a}{2187 \pi  m^3}\,.\nonumber
	\end{align}
The energy-momentum tensor is not fully fixed, but rather depends on a set of free parameters: $q_{2m,2n}$ and the embedding of the shell in the coordinate system outside the shell $x_{2m,0}$. This is just like \cite{Danielsson:2021ruf}, where demanding certain properties of the stress-tensor fixed the quadrupole moment, while leaving the radius at order $a²$ unfixed.

\subsection{Perfect fluid}\label{sec:perfect_fluid}
As a first check of our approach using the Ernst potential, let us follow in the same vein as \cite{Danielsson:2017riq,Danielsson:2017pvl,Danielsson:2021ruf}, and demand that the stress tensor obtained above is a perfect fluid \footnote{As we will see in the next section, it will be necessary to relax this assumption, with very interesting consequences.} \ie
\begin{equation}
	S^a{}_b = ρ u^a u_b + p (δ^a{}_b + u^a u_b)\,,
\end{equation}
where $u^a$ is the velocity vector of the fluid element. This can be easily read off from $S^a{}_b$ by recalling that $u^a$ is an eigenvector with eigenvalue $-ρ$ (\ie projecting along $u^a$ gives $S^a{}_b u^b = - ρ u^a$).
This determines $q_{2,2}$
\begin{equation}
	q_{2,2} = \frac{2048}{243 m² \left( 261 \ln 3 - 196 \right)} + \mathcal{O}\left( m^{-3}k^{-1}\right)\,.
\end{equation}
Using \eqref{eq:multipoles}, we see that this uniquely determines the quadrupole moment at order $a²$. At leading order in $a$ and $k$,
	\begin{align}
		M_0 &= m - a²m q_{2,2}\nonumber\\
		& = m-\frac{2048 a^2}{243 m (261 \ln 3-196)}\nonumber\\
		& = m -\frac{0.0928828 a^2}{m} +\mathcal{O}\left( 1/k \right)\,,\nonumber\\
		J_1 &= a m\,,\\
		M_2 &= -a²m + \frac{4}{5}a²m³ q_{2,2}\nonumber\\
		& = -a^2 m + \frac{8192 a²m}{1215 (261 \ln 3-196)}\nonumber\\
		& = -0.925694 a²m + \mathcal{O}\left( a²/k \right)\,.\nonumber
	\end{align}
To identify the metric parameters $m$ and $a$ with physical mass and angular momentum respectively, we redefine
\begin{equation}
	m = \frac{2048 A^2}{243 M (261 \ln 3-196)}+M,\quad a = A\,,
\end{equation}
which gives
\begin{equation}\label{eq:quad_perfect}
	M_0 = M,\quad J_1 = A M,\quad M_2 =-0.925694  A²M\,.
\end{equation}
It is satisfying that the quadrupole thus obtained is precisely the one that was obtained in \cite{Danielsson:2021ruf} using a very different analysis.
Using \eqref{eq:paramout} and \eqref{eq:jc1_sol_x}, this fixes the shape of the shell, but leaves the size at order $a²$ arbitrary.
\begin{align}\label{eq:xout_perfect}
	x &= \frac{5}{4} + a² \left( x_1 - \frac{2 y^2 (52+495 \ln 3)}{81 m^2 (261 \ln 3-196)}  \right)\nonumber\\
	& = 1.25 + a² \left( x_1 - \frac{0.162131 y²}{m²} + \mathcal{O}\left( k^{-1}m^{-1} \right) \right) \,.
\end{align}
This situation is completely analogous to the result obtained in \cite{Danielsson:2021ruf}. This perfect fluid on the shell can be decomposed into pieces corresponding to tension, radiation, and stiff matter. 
\begin{equation}
    S_{\textrm{full}} = S_{\textrm{gas}} + S_{\textrm{brane}} + S_{\textrm{stiff}}\,.
\end{equation}
At leading order in spin (i.e. $a^2$) and $k$, energy-momentum tensor of the traceless radiation with equation of state $ρ=p/2$ is given by \eqref{eq:a2traceless}, while the non-zero elements of the piece corresponding to the brane tension with equation of state $ρ=-p$ is given in  \eqref{eq:tension}.
\begin{widetext}
\begin{equation} \label{eq:a2traceless}
\begin{aligned}
    \left(S_{\textrm{gas}}\right)^t{}_t &= -\frac{1}{27 \pi } + \frac{a^2 \left(1944 x_1 (261 \ln 3-196)+128 y^2 (1071 \ln 3-1228)+170620-208395 \ln 3\right)}{13122 \pi  (261 \ln 3-196)},\\
    \left(S_{\textrm{gas}}\right)^θ{}_θ &=\frac{1}{54 \pi }+\frac{a^2 \left(-1944 x_1 (261 \ln 3-196)+128 y^2 (52+495 \ln 3)-20092+7947 \ln 3\right)}{26244 \pi  (261 \ln 3-196)},\\
    \left(S_{\textrm{gas}}\right)^ϕ{}_ϕ &= \left(S_{\textrm{gas}}\right)^θ{}_θ -\frac{128 a^2 \left(y^2-1\right)}{2187 \pi  m^3},
    \left(S_{\textrm{gas}}\right)^t{}_ϕ = \frac{2 a \left(y^2-1\right)}{9 \pi }, 
    \left(S_{\textrm{gas}}\right)^ϕ{}_t =\frac{32 a}{2187 \pi }\,,
\end{aligned}
\end{equation}
\begin{equation} \label{eq:tension}
\begin{aligned}
    \left(S_{\textrm{tension}}\right)^t{}_t &= 
    \left(S_{\textrm{tension}}\right)^θ{}_θ = 
    \left(S_{\textrm{tension}}\right)^ϕ{}_ϕ =\\
     &-\frac{k}{4 \pi }+\frac{2}{27 \pi } -\frac{a^2 \left(648 x_1 (261 \ln 3-196)+32 y^2 (765 \ln 3-292)+17492-32697 \ln 3\right)}{6561 \pi  (261 \ln 3-196)},
\end{aligned}
\end{equation}
\end{widetext}

The stiff matter part of $S^a{}_b$ starts at order $1/k$, and is therefore sub-leading to the order that we are working.
Repeating the analysis in \cite{Danielsson:2021ruf}, we also find the same values for the flux coefficients $α,β$, completing a highly non-trivial consistency check of our formulation.

We can now extend the analysis of the previous sections to order $a^6$. Requiring that the stress-tensor is a perfect fluid to that order determines the multipolar parameters $q_{m,n}$ whose numerical values are listed below.
\begin{align}\label{eq:q_prefect}
	q_{2,2} &= \frac{0.0928828}{m^2}\,, \nonumber\\
	q_{2,4} &= \frac{0.339375}{m²} x_1 + \frac{0.0704772}{m^4}\,,\nonumber\\
	q_{2,6} &= \frac{0.379561}{m^4} x_1-\frac{0.0729619}{m^6}\,,\nonumber\\
	q_{4,4} &= -\frac{0.0547387}{m^4}\,,\\
	q_{4,6} &= -\frac{0.385026}{m^4} x_1 + \frac{0.0117411}{m^6}\,,\nonumber\\
	q_{6,6} &= \frac{0.0790248}{m^6}\,.\nonumber
\end{align}
The full expressions for these, as well as the analysis of this section repeated to order $a^6$ can be found in the Wolfram Mathematica notebook \cite{mathematica:a6}.
	
\subsection{Traceless fluid}\label{sec:traceless_fluid}
It is tempting to stop here and declare victory. 
We have reproduced our lowest order results from before, obtained new results to sub-sub-leading order in spin, and provided a prescription to construct arbitrarily fast spinning black shells. However, it is clearly unsatisfactory that the radius remains unfixed at higher orders in spin. Could there be a physical principle that we have missed?
	
Let us pause and look back at the original construction of a static black shell in \cite{Danielsson:2017riq}. If one only considers the black shell after it has been formed, its radius remains a free parameter. What pins the shell to its Buchdahl radius is further consideration of the physics of the problem. A crucial input to the model is that the degrees of freedom of a collapsing shell of matter are converted to radiation (traceless fluid) right on top of the shell. The easiest way to account for this is to consider energy-momentum of the Schwarzschild bubble that is in excess of a vacuum Minkowski solution. This was identified with $ρ_b, p_b$ in \cite{Danielsson:2017riq} and gave the Buchdahl radius. Consistency of the analysis was further confirmed by the considering the entropy and thermodynamics of the bubble in that article.
	
For the rotating black shell that we are now considering, it is natural to follow the same reasoning and require that the stress tensor associated to the infalling matter is traceless. This is obtained by performing a subtraction with empty space, as discussed at the beginning of \fref{sec:rotating}. Although tracelessness was commensurate with $S^a{}_b$ being a perfect fluid for the stationary black shell in \cite{Danielsson:2017riq}, this is no longer the case even to lowest order in spin. So we will now impose tracelessness {\it without} requiring the stress tensor to be a perfect fluid.
This can be done to arbitrary order in $a$ and fixes all the remaining free parameters, resulting in a shell with a unique size and shape. To order $a^6$, these are (presenting only numerical values here for brevity)
\begin{equation}\label{eq:q_traceless}
	\begin{aligned}
	q_{2,2} = -\frac{0.045}{m^2}, 
	q_{2,4} = -\frac{0.070}{m^4}, 
	q_{2,6} = -\frac{0.178}{m^6},\\
	q_{4,4} = \frac{0.408}{m^4},
	q_{4,6} = -\frac{0.008}{m^6},
	q_{6,6} = -\frac{0.327}{m^6},\\
	x_{2,0} = \frac{0.332}{m^2},
	x_{4,0} = -\frac{0.107}{m^4},
	x_{6,0} = \frac{0.808}{m^6}\,.
\end{aligned}
\end{equation}
Note that the parameters above differ from those in \eqref{eq:q_prefect} that were obtained for a perfect fluid on the shell. This fixes the multipolar structure of the metric completely. Redefining the metric parameters $m$ and $a$ to coincide with the physical mass and angular momentum as before gives (for large $k$) yields
\begin{align}\label{eq:multipole_traceless}
	M_0 &= M,\quad
	M_2 =-1.036 A^2 M +\frac{0.271 A^4}{M} -\frac{0.719 A^6}{M^3},\nonumber\\
    M_4 &= 1.142 A^4 M-\frac{0.968 A^6}{M},
    M_6 = -1.309 A^6 M,\\
    J_1 &= A M,
	J_3 = \frac{0.544 A^5}{M}-1.073 A^3 M,\nonumber\\
	J_5 &= 1.213 A^5 M\,.\nonumber
\end{align}
The shell, as seen from outside, now sits at
	\begin{align}\label{eq:radius}
		x &= 1.25 +\frac{a²}{m²} \left(0.332-0.325 y^2\right)\nonumber\\
		&+ \frac{a^4}{m^4} \left(-0.108-0.229 y^2+0.436 y^4\right)\\
		&+ \frac{a^6}{m^6} \left(0.808-0.403 y^2-0.113 y^4-0.510 y^6\right)\nonumber\\
		& + \mathcal{O}\left(a^8/m^8\right)\,.\nonumber
	\end{align}
\begin{figure}
    \centering
    \includegraphics[width=\linewidth]{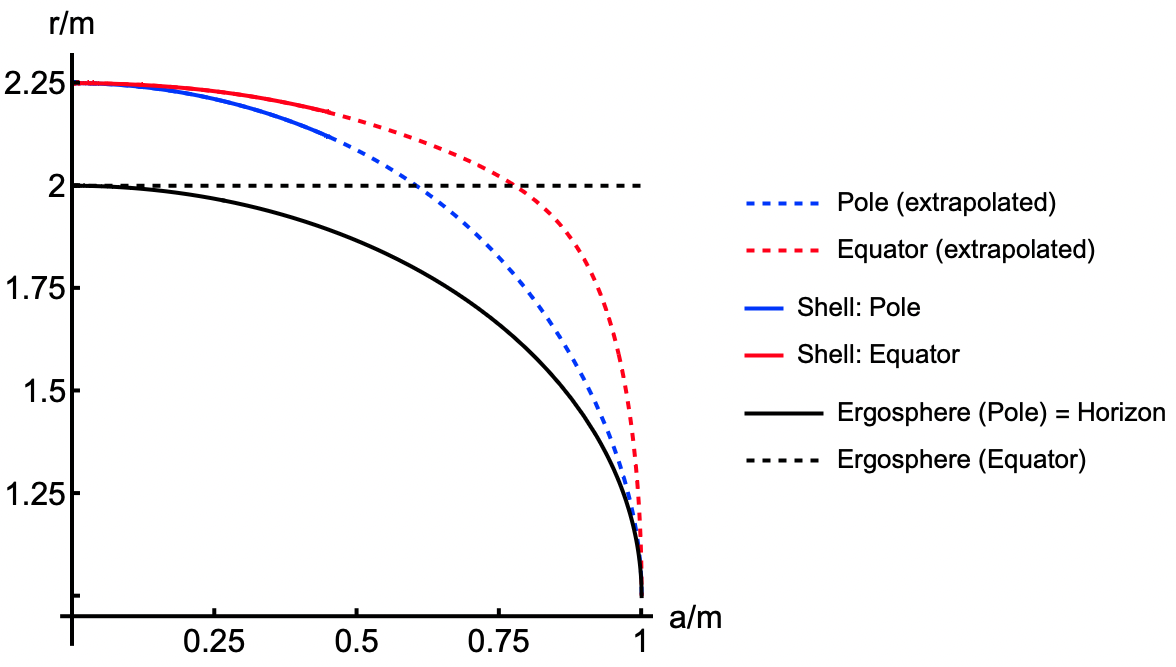}
    \caption{The polar and equatorial radii of the spinning shell as a function of its spin are shown with blue and red lines respectively. Radius of the horizon and ergosphere are shown with black lines. Since the deviation from the Kerr geometry is small, these are very close to the Kerr values. We have computed explicitly to order $a^6$, and our results are accurate for $0 \leq a \leq 0.45m$. As a speculative estimate, we have extrapolated the functional form of the radii for $0.45m \leq a<m$. Curiously, this indicates that the shell asymptotes to the horizon in the extremal limit, just like it does for extremal Reissner-Nordström in \cite{Danielsson:2017riq}.}
    \label{fig:radius}
\end{figure}
This shows that the shell starts to shrink and gets more and more oblate as it rotates faster. The polar and equatorial radii of the shell are shown with solid blue and red lines in \fref{fig:radius}. While we can make accurate predictions only to order $a^6$, we cannot help but speculate about the ultimate fate of the shell as it approaches extremality. Since $\sigma \rightarrow 0$ as $a \rightarrow m$, the multipole moments given in (\ref{eq:multipoles}) suggest that the metric approaches that of extremal Kerr. By naively extrapolating the radius as $a→m$, the shell seems to approach the horizon in this limit. Curiously enough, this is exactly what was found for an extremal Reissner-Nordström shell in \cite{Danielsson:2017riq}. The exterior metric, being very close to Kerr turns out to have an ergosphere that is almost identical to that of a Kerr black hole. This is shown with black lines in \fref{fig:radius}. Let us now make some speculative comments based on the extrapolation above. Around $a\sim 0.6m$, the shell seems to shrink beyond its ergosphere, and one would expect to see a sliver of the ergosphere peeping out around the equator, with more of it showing as the shell approaches extremality. At extremality, the shell seems to sit very close to the Kerr horizon, with a fully formed ergosphere.\footnote{Since the location of the Kerr horizon is actually a naked singularity for the QM metric, it would be interesting to study the physics as the shell approaches this location.}
It was argued in \cite{Friedman-ergoregion} that a compact object with an ergoregion but without a horizon suffers from instabilities. However, \cite{Maggio:2017ivp} showed that the role of the horizon in removing the instability is by effectively absorbing incoming negative energy states. They show that a compact object with an absorption coefficient as small as $1\%$ can avert this instability. Black shells are expected to be excellent absorbers; it is therefore reasonable to expect that they are immune to an ergoregion instability. 
Since the ergoregion is expected to appear around $a\sim 0.6m$, one would need to go to at least order $a^{10}$ to do a concrete computation. This is a challenging computation that is beyond the scope of the present paper, but certainly something that we would like to return to in the future.

To summarize this section, we have used an additional physical input: the rotating shell of matter that collapses to form the black shell contributes a traceless fluid on top of the shell. This determines the shape and size of the shell, and fixes the multipolar structure of the spacetime outside a black shell.
Having done the analysis to order $a^6$, the above expressions are accurate to $a^6 \sim 1\% ⇒ a \sim 0.45m$. This predicts percentage level deviation in multipole moments with respect to Kerr (where $δM_n \coloneqq (M_n/M_{\textrm Kerr})-1$)
\begin{equation}
	δM_2 \sim 1.13\%,\quad 
	δM_4 \sim -6.65\%,\quad
	δM_6 \sim 30.8\%\,.
\end{equation}
%
%
The quadrupole moment obtained here is different from the one obtained using a perfect fluid in \eqref{eq:quad_perfect} and \cite{Danielsson:2021ruf}, since two results are obtained using different arguments. Both arguments are mathematically consistent, but given the physics of the problem, we believe that requiring the collapsing matter to form a traceless fluid on top of the shell is staying true to the spirit of the original spherically symmetric configuration in \cite{Danielsson:2017riq}. In \fref{sec:stability} we will elaborate on the physics behind this condition and argue for a thermodynamic explanation for it, at least for the non-rotating case. Let us now explore the far reaching consequence of the black shell fluid not being perfect. This will involve the full power of relativistic hydrodynamics.

\section{Relativistic hydrodynamics of a rotating black shell}\label{sec:hydrodynamics}
	
In section \ref{sec:traceless_fluid} we obtained the fully fixed energy momentum tensor $S^a{}_b$ of the fluid supporting the black shell. To understand the physical properties of this fluid, we will resort to relativistic hydrodynamics. This is challenging, given that the subject is still not fully developed.
	
It is easy to check that the explicit $S^a{}_b$ that we have found can not be a perfect fluid.
This is not a surprise. For instance, as discussed in \cite{Bemfica:2020zjp} it is a very general result that the shear of a relativistic fluid is related to its capability to absorb gravitational waves. For systems involving black branes, you typically expect the shear to be proportional to the entropy density. The total absorption cross section of the black shell is proportional to its area, as is the total entropy which is of order $R^2/\ell_4^2$ (where $R$ is the radius of the shell, and $\ell_4$ is the 4d Planck length). The entropy density goes as $s \sim 1/\ell_4^2$, and we expect the shear to be of a similar size $\eta \sim 1/\ell_4^2$.

For a relativistic fluid system, there are two conserved quantities: the particle current $J^a$ and the energy-momentum tensor $S^{ab}$:
\begin{equation}\label{eq:conservation}
    ∇_a S^a{}_b = 0 = ∇_a J^a\,.
\end{equation}
The most general way of decomposing the energy-momentum tensor  in terms of a time-like unit vector $u^a$ (\ie\, $u^a u_a = -1$) is as follows:
\begin{equation}\label{eq:Sgeneral}
	S^a{}_b = \mathcal{E}u^a u_b + \mathcal{P} Δ^a{}_b +u^a \mathcal{Q}_b + u_b \mathcal{Q}^a + \mathcal{T}^a{}_b\,,
\end{equation}
where $Δ^a{}_b \coloneqq δ^a{}_b + u^a u_b$. The scalars defined above are $\mathcal{E} \equiv S^a{}_b u_a u^b$, $\mathcal{P} \equiv Δ_a{}^b S^a{}_b/2$, the vector is $\mathcal{Q}^a \equiv -Δ_c{}^a S^c{}_d u^d$, and the symmetric traceless tensor is 
$\mathcal{T}^{ab} \equiv Δ^{ab}{}_{cd}S^{cd}$, where $ Δ^{ab}{}_{cd} \coloneqq (1/2) \left( Δ^a{}_cΔ^b{}_{d} + Δ^a{}_d Δ^b{}_c -Δ^{ab}{}\Delta _{cd}\right)$
Physically, $\mathcal{E}, \mathcal{P}, \mathcal{Q}$, and $\mathcal{T}^a_b$ are (out of equilibrium) energy density, pressure, heat flow and the shear respectively.
On the other hand, the particle number flow $J^a$ can be decomposed in terms of $u^a$ as
\begin{equation}\label{eq:j}
	J^a = \mathcal{N} u^a + {\cal J}^a\,,
\end{equation}
where $\mathcal{N}$ is the number density of particles in the fluid, while $\mathcal{J}^a$ is an additional current. In the absence of the second term, conservation of the particle current implies particle number conservation. But for non-zero $\mathcal{J}^a$, the change in particle number can be compensated by this flow vector.

At equilibrium, a fluid can be parametrized by temperature $T_{\textrm eq}$, flow velocity $u^a_{\textrm eq}$, and the chemical potential $μ_{\textrm eq}$. Out of equilibrium, these quantities $T,u^a,μ$ are not well defined. Different out-of-equilibrium values of these quantities are allowed as long as their equilibrium values agree. A particular choice of how one defines $T,u^a,μ$ in an out-of-equilibrium fluid is called a choice of \emph{hydrodynamic frame} or simply \emph{frame}. A change of frame is just a field redefinition of $T,u^a,μ$ by derivative corrections. Viscous hydrodynamics is studied as a derivative expansion, with gradients of $T, u^a, μ$ parametrizing deviation from equilibrium. 

\begin{figure*}
    \centering
    \subfigure[$u^t$]{\includegraphics[width=0.23\linewidth]{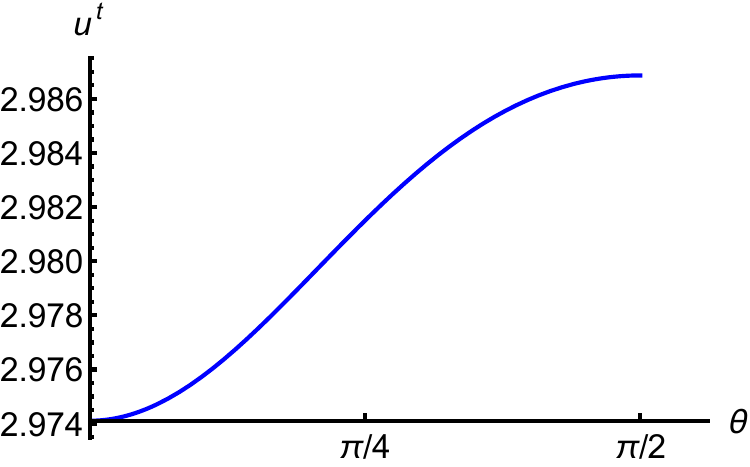}}
    \hfill
    \subfigure[$u^y$]{\includegraphics[width=0.23\linewidth]{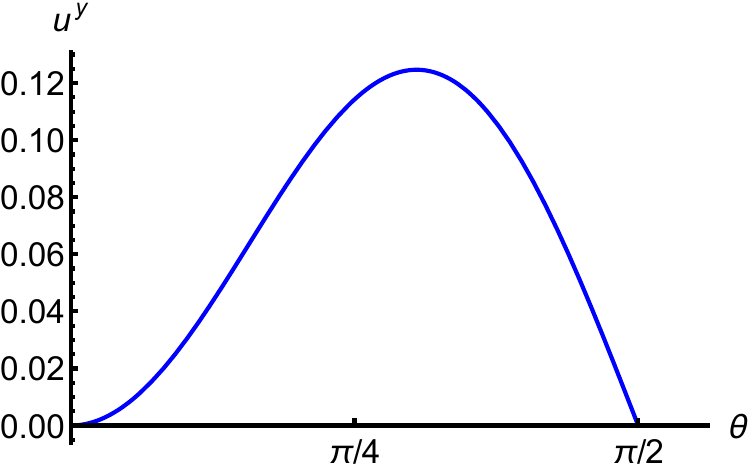}}
    \hfill
    \subfigure[$u^ϕ$]{\includegraphics[width=0.23\linewidth]{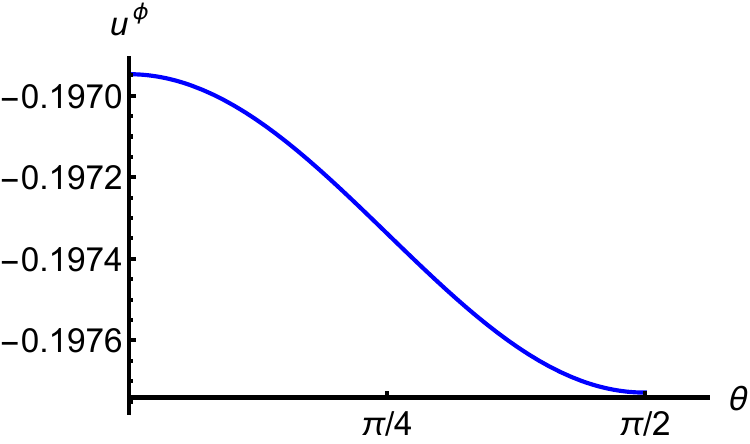}}
    \hfill
    \subfigure[Unruh temperature]{\includegraphics[width=0.23\linewidth]{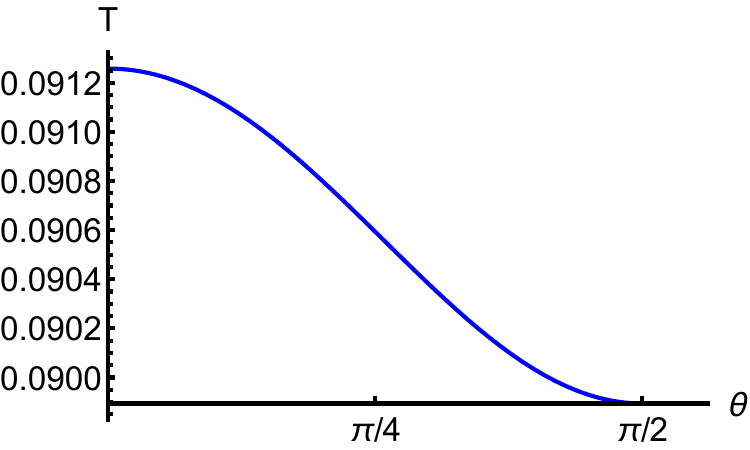}\label{fig:unruh}}
    \caption{(a),(b),(c) show components of the velocity vector of the radiation fluid as a function of the polar angle $θ$. (d) shows the Unruh temperature in the frame of the fluid for $a=0.3, m=1$. As a function of $a$, and $y$, this is given by $T = 0.094 + a^2 \left(0.011 y^2-0.055\right) + a^4 \left(-0.0083 y^4+0.057 y^2+0.066\right) + \mathcal{O}\left(a^6\right)$.
    }
    \label{fig:ua}
\end{figure*}

Historically, there are two famous frame choices: (i) Eckart's frame, where one chooses $\mathcal{E}=ε, \mathcal{N}=n, \mathcal{J}^a = 0 $; (ii) Landau-Lifschitz's frame, where one chooses instead $\mathcal{E}=ε, \mathcal{N}=n, \mathcal{Q}^a = 0$. Here $ε$ and $n$ are the equilibrium energy and number of particles respectively. 
However, more recently, it has been realized that neither of these frames result in a causal and well defined theory. 
A first order gradient expansion that includes first derivatives of the temperature, velocity, and chemical potential in all quantities \eqref{eq:Sgeneral} and \eqref{eq:j} was proposed by BDNK \cite{Bemfica:2020zjp,Kovtun:2019hdm}. This theory is causal and hyperbolic.
%
%
\begin{align}\label{eq:grad_expansion}
    \mathcal{E} &= ε + ε_1 u^a  \frac{∇_a T}{T} + ε_2 ∇_a u^a + ε_3 u^a ∇_a\left(μ/T\right)\,,\nonumber\\
    \mathcal{P} &= P + π_1 u^a  \frac{∇_a T}{T} + π_2 ∇_a u^a + π_3 u^a ∇_a\left(μ/T\right)\,,\nonumber\\
    \mathcal{N} &= n + ν_1 u^a  \frac{∇_a T}{T} + ν_2 ∇_a u^a + ν_3 u^a ∇_a\left(μ/T\right)\,,\\
    \mathcal{Q}^a &= θ_1 Δ^{ab} \frac{∇_b T}{T} + θ_2 u^b ∇_b u^a + θ_3 Δ^{ab}∇_b\left(μ/T\right)\,,\nonumber\\
    \mathcal{J}^a &= γ_1 Δ^{ab} \frac{∇_b T}{T} + γ_2 u^b ∇_b u^a + γ_3 Δ^{ab}∇_b\left(μ/T\right)\,,\nonumber\\
    \mathcal{T}^{ab} &= -2η Δ^{abcd}∇_c u_d\,.\nonumber
\end{align}
The equilibrium state characterized by $ε, P, n$ above corresponds to the static shell, which is a perfect fluid. The quantities $\mathcal{E}, \mathcal{P}, \mathcal{Q}$, and $\mathcal{T}^a_b$, appearing in the energy momentum tensor, are all of order $1/(\ell_4^2 R)$, compatible with the mass of the black shell given by $M \sim \epsilon \times R^2 \sim R/\ell_4^2$. Since $T$, $u^a$ and $\mu$ vary over the size of the shell, we have $\nabla _a \sim 1/R$, from which it follows that the scales of the parameters $\epsilon_i$, $\pi_i$, $\theta_i$, $\gamma_i$, and $\eta$ are set by $1/\ell_4^2$. To further understand the system, it is also necessary to make an expansion in powers of derivatives, including contributions such as $(\nabla u)^2$ and $\nabla^2 u$. The zeroth order case is the perfect fluid, while keeping first order in the derivatives as we did above correspond to a viscous fluid. Keeping higher derivatives corresponds to $k$-th order thermodynamics. By dimensional analysis such terms will be suppressed by powers of $l/R$, where $l$ is a scale determined by the microscopic physics of the fluid. It is not important for us whether this scale is set by the Planck scale $\ell_4$, or the AdS-scale of the interior. In any case, these corrections will be small for a macroscopic black hole and we can expect the viscous fluid to describe the rotating black shell for any value of the angular momentum.
%
	\begin{figure}
     \centering
     \subfigure[Matter flow]{\includegraphics[width=0.45\linewidth]{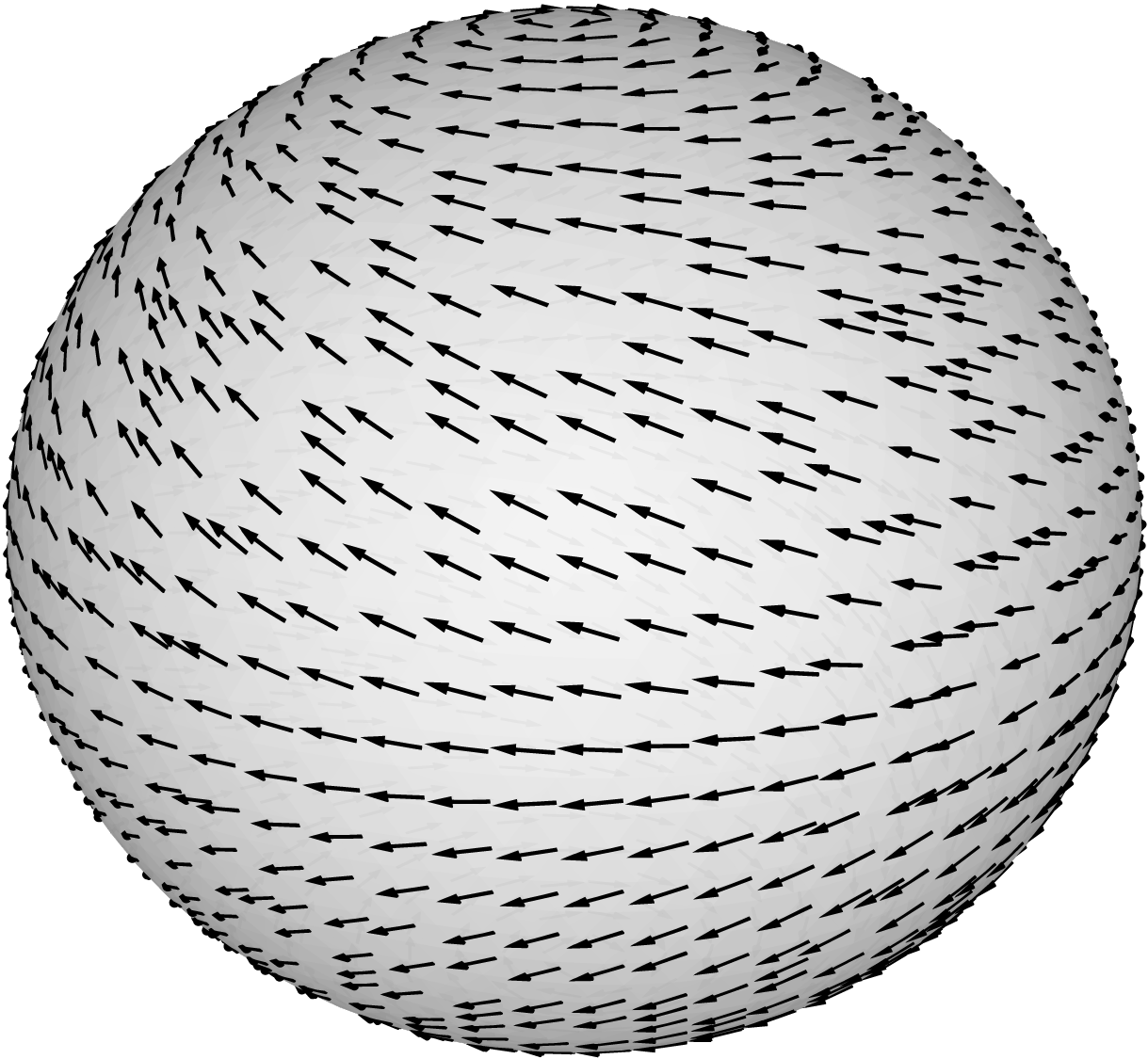}\label{fig:matter_flow}}
     \hfill
     \subfigure[Heat flow]{\includegraphics[width=0.45\linewidth]{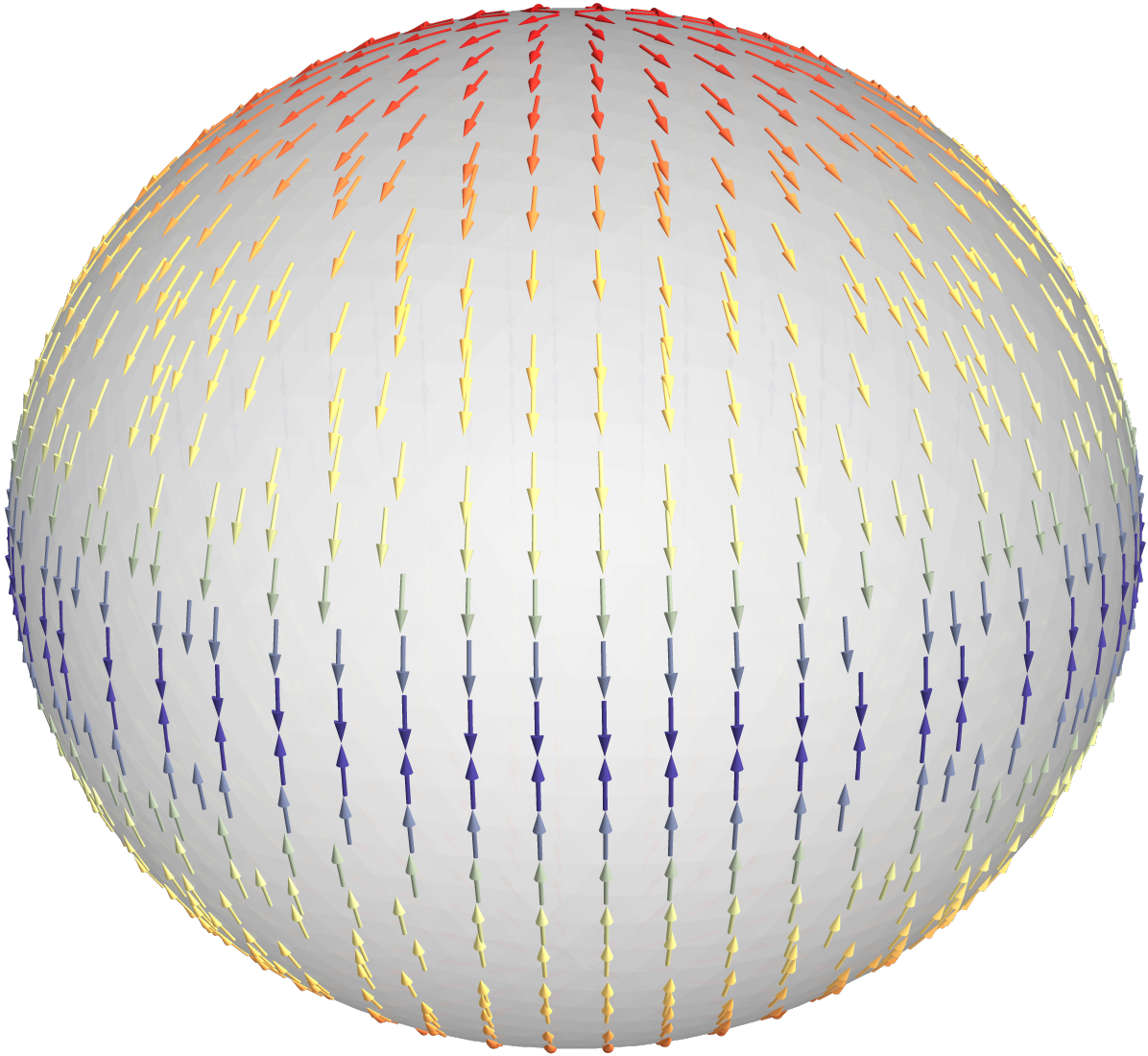}\label{fig:heat_flow}}
    \caption{Velocity vector of the traceless fluid on a rotating black shell. Matter flows from the equator to the poles, with a rotational component due to spin. The shell is oblate, so the poles have higher Unruh temperature than the equator. Heat flows from the poles (hotter) to the equator (cooler). 
    }
    \label{fig:matter_heat_flow}
\end{figure}
%
Before trying to match the energy momentum tensor to this form, let us figure out what to expect. The temperature of the fluid will be given by the Unruh temperature, and as we will see in \fref{fig:unruh}, the poles have a higher Unruh temperature than the equator. The resulting temperature gradient will drive a flow of heat from the poles towards the equator. Given that we have a stationary situation with a net energy flow only in the direction of rotation, this implies that the fluid must be flowing towards the poles to compensate. There is, then, a circulating current of energy flux towards the poles in form of the fluid and away from the poles as heat. The circulation is sustained by the Unruh effect that makes sure that the difference in temperature remains despite the flow of heat. Note that the energy as well as the entropy remains constant over a very long time. It is only due to the loss of radiation to the bulk, at time scales of order of Hawking evaporation, that the system eventually dissipates. Amusingly, this reminds us of the trade winds on the rotating Earth, except that the flow is reversed. The trade winds blow towards the equator, where the air masses are heated up, rise, and flow back towards the poles at high altitudes. This is very much like what the fluid does on the shell, except that it flows towards the poles, and the heat flows back to the equator. Just as the sun is the source of heat for the weather systems on the Earth, it is the Unruh effect that drives the circulation on the black shell. See figure \ref{fig:matter_heat_flow}.

An interesting aspect of relativistic fluid flow is that, as a consequence of \eqref{eq:conservation}, the combined effect of fluid flow and the heat flow is conserved, rather than the number density of the fluid itself being conserved. As a consequence, we need ${\cal J}^a \neq 0$ and cannot choose Eckart's frame.\footnote{It was observed in \cite{Kovtun:2019hdm} that Eckart's frame (where $\mathcal{J}^a=0$) is only compatible with a conserved particle number. Such fluids are often called {\it charged}, and the conservation of $J^a$ is the same as conservation of charge.} ${\cal J}^a$ is often called the conductive particle current, and compensates for the non-conserved particle number to retain a conserved current. The result is that the conserved current $J^a$ will point only in the direction of rotation without any latitudinal component. Since $S^{ty}=S^{\phi y}=0$, any observer not moving in the $y$-direction, will observe an energy flux $S^{0\mu}$ in the direction of the rotation of the shell, with $S^{0y}=0$. Since $u^y \neq 0$, and ${\cal Q}^y \neq 0$, the observer concludes that the flow of heat compensates for the flow of the fluid such that there is no net energy flux along the latitudes.  

Let us now see how this works out in more detail. Unlike the case of a perfect fluid, $u^a$ is no longer an eigenvector of $S^a{}_b$, but is instead determined so that the shear piece of the explicit $S^a{}_b$ we obtained in the previous section, becomes of the form given just below \eqref{eq:Sgeneral}. This fixes $u^a$ up to integration constants, and higher order constants that would be determined at the next order in $a$.
%
%
If we, furthermore, require the full $S^a{}_b$ to be of the form given \eqref{eq:Sgeneral} with the first order gradient expansion in \eqref{eq:grad_expansion}, most of these parameters are fixed, with two remaining free parameters at this order $θ_{12}, θ_{22}$; and we find
\begin{align}
    u^a &= \left\{3+a² f_{1,2}(y)+a^4 f_{1,4}(y) \right.,\nonumber\\
    & \quad a² f_{2,2}(y)+a^4 f_{2,4}(y),\\
    & \quad \left. a f_{3,1}(y) + a³ f_{3,3}(y)\right\}+\mathcal{O}\left(a^5\right)\,,\nonumber
\end{align}
where the functions $f_{i,j}(y)$ are given by
\begin{align}
f_{1,2}(y) &= \frac{1}{m^2}\left( -0.132 y^2-0.255 \right),\nonumber \\
   f_{2,2}(y) &= \frac{1}{m^3}0.273 y (1-y²),\nonumber\\
   f_{3,1}(y) &= -\frac{0.658}{m^2}\,,\nonumber\\
   f_{1,4}(y) &= \frac{1}{m^4} \left( -0.5 y^4+0.388 y^2+1.21 \right),\\
   f_{3,3}(y) &= \frac{1}{m^4}\left( -0.0073 + 0.029 y² \right),\nonumber\\
   f_{2,4}(y) &= \frac{y\left(1-y²\right)}{m^5} \left[ 1.827 -0.539 y^2 \right.\nonumber\\
   & \qquad \left. -m^2 (1.275 θ_{12} + 0.49 θ_{22})\right]\,.\nonumber
\end{align}
The $u^a$ above is plotted in \fref{fig:ua}, and we see that the $y$-component of the fluid velocity vanishes at the poles and on the equator. This is in agreement with our general discussion at the beginning of the section, with the particle number of the fluid not conserved. At the poles particles are dissolved into heat that flows back to the equator (\fref{fig:heat_flow}), while on the equator, heat is converted into fluid that flows back to the poles (\fref{fig:matter_flow}). All is driven by the temperature difference sustained by the Unruh effect (\fref{fig:unruh}).

The gradient expansion determines the shear viscosity $η$, and thermal transport coefficients $θ_i$ to be \footnote{Note that, since these are proportional to derivative terms (which themselves start at order $a^2$), the transport coefficients $η,θ_i,ε_i,π_i$ are only determined to order $a^2$ although we are matching the heat flow and shear to order $a^4$.}
\begin{align}\label{eq:ηθsol}
    η &= 0.0078+ a^2 \left(0.036\, θ_{12} + 0.014\, θ_{22} -\frac{0.072}{m^2}\right) + \mathcal{O}\left(a^4\right),\nonumber\\
    θ_1 &= -0.214 + a^2 θ_{12} + \mathcal{O}\left(a^4\right),\\
    θ_2 &= 0.000712 + a^2 θ_{22} + \mathcal{O}\left(a^4\right).\nonumber
\end{align}
We furthermore find that \eqref{eq:Sgeneral} with \eqref{eq:grad_expansion} for the energy, pressure and particle number implies
\begin{align}
    ε &= \frac{0.023}{m} - 0.0118 \frac{a^2}{m^3}  + 0.022 \frac{a^4}{m^5} + \mathcal{O}\left(a^6\right),\nonumber\\
    ε_1 &= 0.0573 + a^2 ε_{12}+ \mathcal{O}\left(a^4\right),\nonumber\\
    ε_2 &= 0.00139 - a^2 \left(0.00652 θ_{12} + 0.0025 θ_{22} + \frac{0.00567}{m^2}\right)\nonumber\\
    & \quad + \mathcal{O}\left(a^4\right),\nonumber\\
    P &= \frac{0.0118}{m} - 0.0059 \frac{a^2}{m^3} + 0.011 \frac{a^4}{m^5} + \mathcal{O}\left(a^6\right),\nonumber\\
    π_1 &= 0.0286 + a² π_{12} + \mathcal{O}\left(a^4\right),\\
    π_2 &= -0.000697 - a^2 \left(0.00326\, θ_{12} + 0.00125\, θ_{22} + \frac{0.0028}{m^2}\right)\nonumber\\
     & \quad + \mathcal{O}\left(a^4\right),\nonumber\\
    n &= \frac{0.0354}{m} - 0.0192 \frac{a^2}{m^3} + 0.0376 \frac{a^4}{m^5} + \mathcal{O}\left(a^6\right),\nonumber\\
    ν_1 &= -0.06 + \mathcal{O}(a²),\nonumber\\
    ν_2 &= -0.0047 - a^2 \left(0.022\, θ_{12} + 0.0084\, θ_{22} - \frac{0.012}{m^2}\right)\nonumber\\
     & \quad + \mathcal{O}\left(a^4\right)\,.\nonumber
\end{align}
$ε,P,n$ that we have recovered above represent the equilibrium values. Interestingly, we see that in addition to the stationary values of $ε, P$ from \cite{Danielsson:2017riq}, there are additional spin dependent contributions.
The equation of state is $P=ε/2$ corresponding to radiation, as required by traceleness imposed in \fref{sec:traceless_fluid}. Here we have assumed the temperature to be the local Unruh temperature in the frame of the fluid, which is shown in \fref{fig:unruh}. 

Let us look at our results in somewhat more detail, and compare with expectations. In \cite{Kovtun:2019hdm} it was argued that the constants in \eqref{eq:grad_expansion} must satisfy a number of constraints in order for the equilibrium of the system to make sense. In particular, it was argued that the temperature (as well as the chemical potential) should be proportional to $1/\sqrt{g_{tt}}$. It was also argued, based on this, that the divergence of the velocity of the fluid, the heat flow as well as the shear, should vanish at equilibrium. These are all sensible conditions which, among other things, imply that $\theta_1=\theta_2$ and $\gamma_1=\gamma_2$. As we see from \eqref{eq:ηθsol} this constraint is {\it not} satisfied by our fluid. It is easy to see why. 

For the subtraction discussed in section 2, we match the metric induced by the non-trivial bulk metric, to the metric on a (deformed) shell embedded into empty Minkowski space. Making sure this is possible, together with requiring the resulting energy momentum tensor to be traceless, determines the metric of the bulk as well as the induced metric and the embedding. In particular, it is true by construction that $\sqrt{g_{tt}}$ on the brane is constant. This is a very sensible result, implying that the shell deforms such that there are no gravitational gradient across the surface of the shell. That is, the surface is fully relaxed without any mountains. Following the logic of \cite{Kovtun:2019hdm}, one would expect the temperature to be constant and there to be no heat flow. As we have seen, this is not compatible with the form of  $S^a{}_b$ that we have found. We already know why. In ordinary circumstances quantum effects such as the Unruh temperature can be ignored since it is too small. In our case this is no longer possible. Our system has a particle number density close to Planckian, and the temperature is dominated by the Unruh effect. Even though the gravitational potential is constant across the shell, the Unruh effect sets up a temperature gradient even at equilibrium. This means that the rules of the game change, and we in general find $\theta_1\neq\theta_2$ and $\gamma_1\neq\gamma_2$. One can speculate about other physical equilibria, not related to black shells, where the density is lower and the temperature much higher than the Unruh temperature. Since $\theta_i$ and $\gamma_i$ can depend on the temperature and density, we expect them to approach $\theta_1=\theta_2$ and $\gamma_1=\gamma_2$ in this limit.

Let us conclude this section by pointing out an important simplification that we have made in the analysis of this section. We have \emph{not} included the gradient of the chemical potential in this analysis. We have assumed that these terms can be absorbed in a choice of frame or, alternatively, that they can be traded for gradients of temperature and the velocity vector using on-shell relations. As a next step to our current analysis, we would like to study non-linear stability of perturbations of the black shell numerically along the lines of \cite{Danielsson:2021ykm}. There, it will be necessary to choose a good frame where the system is causal and well-posed. We have not addressed these considerations here since it is not important for our present analysis, but it may then become necessary to revisit this point and refine our choice of frame.

\section{Why the black shell is stable}\label{sec:stability}
	
Perhaps the most difficult challenge when constructing a horizon-less alternative to a black hole is its stability. In case of the black shell, we conjectured already in \cite{Danielsson:2017riq} that a non-trivial transfer of energy between the brane with its tension and the radiation sitting on top of it could result in stability. This was further explored in \cite{Danielsson:2021ykm}, where it was shown, in the non-rotating case, that such processes allow a black shell to absorb infalling matter and grow to a new, stable, radius. 
	
In this section we will propose a detailed microscopic model for the brane constituting the shell, and verify that it does stabilize the system. Some ideas in this direction were discussed in \cite{Danielsson:2017riq}, but what we will argue for here will be somewhat different. For this particular analysis, we will limit ourselves to the non-rotating case, but expect that our arguments will go through in a similar way if the black shell is rotating. We will return to this question in more detail in a future work.
	
There are two crucial properties of the black shell that need to be understood before we can address the issue of stability.
\begin{enumerate}[label=(\roman*)]
    \item First, the number of degrees of freedom of the gas must have a physical explanation. 
    \item Second, the tension of the brane must be able to change.
\end{enumerate}
In \cite{Danielsson:2017riq} we assumed a single brane, which nucleated with its tension at its critical value, given by $\tau=k/4\pi$ (where we are working in units such that $G_4=1$). Here, $k$ is the energy scale of the interior AdS. Let us now instead assume that there are $N$ branes that nucleate when the black shell is formed. Each of these branes has critical tension, such that $N \tau=N\Delta k/4\pi$. Here, $\Delta k$ is the difference in AdS-scale between adjacent vacua in a series of ever deeper AdS-vacua. The larger the value of $N$, the more negative is the vacuum energy of the interior AdS.
	
The main difference with $N$ branes, is that it is now natural to associate the number of degrees of freedom of the massless gas with $N^2$. In this way we can answer the first question posed above. In addition, we expect, in the large $N$-limit, an important correction to the total tension of the stack of $N$-branes. Similar to the discussion in \cite{Danielsson:2022lsl}, in the context of the dark bubble construction of a 4d expanding universe with a positive cosmological constant, we expect there to be $1/N$ corrections to the effective tension.\footnote{The connection with the dark bubble model \cite{Banerjee:2018qey} can also be used to suggest interesting values for the parameters of the black shell model. For a brief review, see \cite{Banerjee:2022myh}. For a discussion of black holes on the dark bubble see \cite{Banerjee:2021qei}. In \cite{Danielsson:2022lsl} it was argued that $R_H \sim N_c \ell_4$, $L\sim N_c^{1/2} \ell_4$, $l_s \sim N_c^{1/4} \ell_4$ and $\ell_5 \sim N_c^{-1/6} \ell_4$. Here $R_H$ is the horizon scale, $L$ the AdS-scale of extra dimension, $l_s$ the string scale, and $\ell_5$ the 5d Planck scale. $N_c \sim 10^{60}$ from matching the observed value of the cosmological constant. Note that $\ell_5 \ll \ell_4$ due to the unique way in which the effective 4d theory is realized. Similarly to how a fundamental string ending on the world-brane give rise to a point particle with Planckian mass $L/l_s^2 \sim 1/\ell_4$, a D3-brane ending on the world-brane will give rise to a 2-brane with tension $L/l_s^4 \sim 1/(L \ell_4^2)$. From here we conclude that $\Delta k \sim 1/L$. Furthermore, if we consider the maximum sized black hole (the Nariai black hole) with its horizon coinciding with the cosmological one, we need $N=N_c$. We then find the total tension of the brane to be $N_c\Delta k/\ell_4^2 \sim 1/\ell_5^3 $. We plan to explore these relations in an upcoming work.\label{foot:dark-bubble}} These should {\it reduce} the tension compared to the critical one, facilitating nucleation in line with the Weak Gravity Conjecture. To be precise, if the physics of the stack of branes is dual to that of, e.g., ${\cal N}=4$ SYM, the tension of the branes could be expected to receive a negative shift of order $1/N^2$ (since $N^2$ is counting the number of degrees of freedom in the adjoint). The total tension then becomes
\begin{equation}
	N \tau \rightarrow N \tau \left( 1 - \alpha/N^2 \right)  = N \tau - \alpha \tau/ N\,,
\end{equation}
where $\alpha$ is a constant of order $1$. The shift makes the tension of the stack slightly less than its critical value, allowing the branes to nucleate at a finite radius. When $N$, which governs the number of degrees of freedom in the gas, changes so does the effective tension of the stack of branes.
	
Let us now compare with the junction conditions. For a shell of radius $r$, these are given by
\begin{align}
	N \tau + \rho &= \frac{N\Delta k}{4\pi} - \frac{1}{4 \pi r} + \rho_b\,,\nonumber\\ 
	-N \tau + p &= -\frac{N\Delta k}{4\pi} + \frac{1}{8 \pi r} + p_b\,,
\end{align}
where the first condition is associated with energy density, while the second one is associated with pressure. On the right hand side of the relations we find contributions from extrinsic curvature,
\begin{equation}
	\rho_b = \frac{1}{4 \pi r}\left(1- \sqrt{1-\frac{2M}{r}}\right) = \frac{1}{6πr} ,
\end{equation}
provided that we choose $r=R_{\textrm B}= 9M/4$, where we also we have $p_b = \rho_b/2 $, thus mimicking the equation of state of radiation. On the left hand side, we find the energy density $\rho$ associated with the fluid on top of the brane, which is expected to include massless radiation. This should exactly match the contribution from the subtracted extrinsic curvature, $\rho_b$, on the right. From the argument given above, we now see that there exists an additional component in $\rho$ given by $-\alpha \tau/N$, coming from the reduced tension of the brane. Since we expect that $N \sim R$, where $R$ is the radius of the shell, this has a chance of matching the remaining term of order $-1/(4\pi r)$ on the right. Let us now see how this comes about in detail. 
	
In \cite{Danielsson:2001xe} the thermodynamics of a black brane (even far from extremality) was understood using an argument based on a competition between the amount of energy in the form of brane tension, and the energy in the form of radiation. By maximizing the entropy in the micro-canonical ensemble keeping the energy constant, the physically correct properties of the black brane, including various non-trivial exponents, could be derived. This was achieved for D3-branes in 10d as well as M2- and M5-branes in 11d. Here, the situation is much more subtle, since the tension of the branes is more or less cancelled against the shift in the background. There is no cost, or gain, while creating branes, {\it except} for the $1/N$-correction. Just as in \cite{Danielsson:2001xe}, we assume that the total energy at our disposal is fixed to $E_0$, and that it can be divided between the brane and the gas. We then get (with the order $1$ constants $\beta _1 =4\pi \alpha$ and $\beta _2$)
\begin{align}
	E_0 &= \beta_2 N^2 T^3 R^2 - \beta_1 \tau R^2/N \,,\nonumber\\
	S &= \frac{3}{2} \beta N^2 T^2 R^2\,,
\end{align}
for the total energy and the total entropy. Solving for the temperature from the first equation, we find that the entropy is given by
\begin{align}
	S &=\frac{3}{2} β_2^{1/3} R^{2/3} N^{2/3} \left( E_0 + \beta_1 \tau R^2/N \right) ^{2/3} \nonumber \\
	&=\frac{3}{2}  β_2^{1/3} R^{2/3} \left( E_0 N+ \beta_1 \tau R^2\right) ^{2/3}\,.
\end{align}
Following \cite{Danielsson:2001xe}, the next step is to find the extremum of $S$ in terms of $N$ by demanding $∂S/∂N=0$. The situation now turns out to be a bit more degenerate, and we find that the only way to have such an extremum is to put $E_0=0$. We will see in a moment why this actually does correspond to the (physical) maximum of the entropy. 
	
With this in mind, let us start by rearranging, as in \cite{Danielsson:2017riq}, the junction conditions  as
\begin{align}
	\rho &= - \frac{1}{12 \pi r}  - \frac{1}{6 \pi r} + \rho_b\,, \nonumber\\ 
	p &= -\frac{1}{24 \pi r} + \frac{1}{6 \pi r} + p_b\,.
\end{align}
On the right we have made the unique decomposition of the $-1/(4\pi r)$ term into a piece corresponding to radiation (the first) and tension (the second). Given that $\rho_b = 1/(6πr)$ at equilibrium, we precisely satisfy the condition obtained from our consideration of the entropy. The extra piece with the equation of state of radiation, is a simple consequence of the fact that the brane can no longer behave as pure tension if $N$ is supposed to vary with the radius of the shell.
	
Having made this non-trivial identification, we will now check that the system not only maximizes the entropy but is also thermodynamically stable. We start by considering the case with an imbalance between the two components such that $E_0$ is perturbed to a non-zero value, holding the radius $r$ fixed. Since the rest of the junction conditions remain unchanged, this is only possible for a {\it negative} value of $E_0$. In this case the reduction in energy can be compensated by kinetic energy from letting the brane move. A positive value of $E_0$ is impossible to account for in this way. We note from \cite{Danielsson:2001xe} that a reduction of $E_0$ to negative values, implies a {\it reduction} of the entropy. Hence, thermodynamics will tend to drive the system back to $E_0=0$ by reducing the kinetic component (through viscous forces) and reheating the system. The entropy then increases back to its largest attainable value.
	
Next, let us consider keeping $E_0=0$ but vary $N$ at fixed $r$. This does not change the entropy, since it is independent of $N$ if $E_0=0$. The invariance corresponds to a scaling of $N$ together with a compensating scaling of $T$ such that $S \sim N^2 T^2 \sim {\rm constant}$. This, on the other hand will shift the temperature away from its Unruh value. Letting the system relax back to its equilibrium implies that the temperature adjusts to the Unruh  temperature at the same time as the number of branes, $N$, changes. Perturbing $r$ off its equilibrium value can be mapped onto changing $N$, so the conclusion is the same also in this case.
	
To summarize, we see that the system is stable against perturbations around the equilibrium. It is interesting to note that there are two different processes at work. One where the Unruh effect restores equilibrium for iso-entropic perturbations. And one where viscous forces drives the entropy back to its maximum. These should match the effective source terms that were used in the phenomenological approach of \cite{Danielsson:2021ykm}.
	
\section{Conclusions and outlook}
	
In this paper we have improved and extended the analysis of rotating black shells in \cite{Danielsson:2021ruf}. We analyze the system to sub-sub-sub-leading order in spin: to order $a^6$. The results we have obtained yield predictions that can be used to distinguish a black shell from a black hole. The shift in the quadrupole starts at order $a^2$, and the corrections we omit are at order $a^8$. An accuracy of $1\%$ at order $a^6$ implies that we can get accurate results to $a = 0.45m$. At this spin, we find an enhancement of the quadrupole moment of $\sim 1\%$. For higher accuracy, one can simply calculate higher corrections.\footnote{While our analysis suggests that the metric approaches Kerr in the extremal limit, we still need to verify what the final fate of the shell will be. We plan to address this in upcoming work.} In an upcoming article we will discuss observations that could be used to detect black shells. While the shift in the quadropole moment is small, it may still be measurable through the tracking of lower mass black holes orbiting supermassive ones. The LISA-observatory is expected to reach the necessary sensitivity.
	
An immediate, and pertinent next step is to test these shells for non-linear stability against perturbations along the lines of \cite{Danielsson:2021ykm}. Since we are no longer limited by spherical symmetry, we can examine stability both in the case of asymmetric accretion of matter where the angular momentum of the shell changes, as well as perturbations with gravitational waves.

The structure with the circulating flows of fluid and heat is extremely intriguing. In its form it reminds of out of equilibrium processes sustained by temperature gradients. This is also what we have here with the Unruh effect acting as the heat source. A difference from the familiar case is that the total system, the gas together with the brane, is to a high degree of accuracy a closed rather than an open system. The heat generated by, say, shear viscosity, is used to create new fluid while keeping the entropy constant and the system in a stationary state. It is only over time scales of the order Hawing evaporation time that the system experiences any real change. As radiation is lost to the bulk, the system decreases in size, even though the total entropy of the black shell and the released radiation increases. This rich and unusual example of relativistic hydrodynamics deserves further studies. None the least, since the subject of relativistic hydrodynamics is in general not well understood.

Another important topic for further studies, is the stringy mechanism behind the transfer of energy between the gas and the brane. We have given some clues to the how and the why of this process in \fref{sec:stability}, which we hope to extend to more general situations such as rotating and colliding black shells. We would also like to explore the intriguing possibility, briefly touched upon in footnote \ref{foot:dark-bubble}, that the black shell can be naturally incorporated into the dark bubble model of de Sitter cosmology.

\addtocontents{toc}{\string\tocdepth@munge}
\section{Acknowledgments}
We are grateful to Alejandro Cárdenas--Avendaño, Luis Lehner, and Frans Pretorius for useful discussions. The work of SG was conducted with funding awarded by the Swedish Research Council grant VR 2022-06157.

\bibliography{refs.bib}
\end{document}